\documentclass[aip,reprint]{revtex4-1}
\usepackage{graphicx}
\usepackage{subcaption}
\usepackage{xcolor}
\usepackage{amsmath}
\begin{document}


\title{Non-orthogonal determinants in multi-Slater-Jastrow trial wave functions for fixed-node diffusion Monte Carlo} 



\author{Shivesh Pathak}
\email[]{sapatha2@illinois.edu}

\author{Lucas K. Wagner}
\email[]{lkwagner@illinois.edu}
\affiliation{Department of Physics; University of Illinois at Urbana-Champaign}


\date{\today}

\begin{abstract}
The accuracy and efficiency of ab-initio quantum Monte Carlo (QMC) algorithms benefits greatly from compact variational trial wave functions that accurately reproduce ground state properties of a system. 
We investigate the possibility of using multi-Slater-Jastrow trial wave functions with non-orthogonal determinants by optimizing identical single particle orbitals independently in separate determinants. 
As a test case, we compute variational and fixed-node diffusion Monte Carlo (FN-DMC) energies of a C$_2$ molecule. 
For a given multi-determinant expansion, we find that this non-orthogonal orbital optimization results in a consistent improvement in the variational energy and the FN-DMC energy on the order of a few tenths of an eV. 
Our calculations indicate that trial wave functions with non-orthogonal determinants can improve computed energies in a QMC calculation when compared to their orthogonal counterparts. 
\end{abstract}

\maketitle 

\section{Introduction}

{A major theoretical challenge in condensed matter physics is the accurate calculation of ground state properties of strongly correlated many-body quantum systems.} 
Finite size systems of interest include single transition metal atoms \cite{BUENDIA201312, doi:10.1063/1.2011393, doi:10.1063/1.4903985}, transition metal oxide molecules \cite{Wagner2003, Wagner2007, PhysRevB.73.075103}, 3d transition metal containing molecules \cite{Dier2016, doi:10.1063/1.2920480}, and transition metal clusters \cite{Sharma2014} for which explicitly correlated many-body methods are necessary to get accurate ground state energies and properties like dipole moments. 
\textit{Ab initio} fixed-node diffusion Monte Carlo (FN-DMC) calculations on these systems have been successful in treating strong correlations and in producing accurate results for understanding low-energy properties. 
However, the accuracy of these FN-DMC calculations are predicated on the ability to choose a trial wave function that accurately reproduces the ground state nodal structure. \cite{Foulkes2001}

{Constructing trial wave functions for FN-DMC that accurately reproduce the ground state nodal structure is extremely challenging due to the high dimensionality of the many-body Hilbert space.}
Compact parameterizations of many-body states can be used to generate trial wave functions that accurately, but still approximately, reproduce the ground state nodal structure.
A common choice of the trial wave function for \textit{ab initio} FN-DMC calculations is the (multi-)Slater-Jastrow ((M)SJ) form. \cite{Foulkes2001}
Since the nodal structure is solely determined by the (multi-)Slater part of a (M)SJ trial wave function, improved nodal surfaces can be achieved by optimizing the single particle orbitals within the multiple Slater determinants.

There have been a number of ways of improving the trial wave function for QMC in the literature. 
Among these, Wagner and Mitas show that the accuracy of calculations for ground state energies and dipole moments of transition metal monoxides in FN-DMC can be increased by using single particle orbitals from a DFT calculation with the B3LYP functional over those from a ROHF calculation. \cite{Wagner2007} 
The orbitals from the DFT calculation can be thought of as relaxed, or optimized, versions of the ROHF orbitals. 
Toulouse and Umrigar describe an efficient parameterization for orbital optimization using orbital rotations\cite{Toulouse2007, Umrigar2007, Toulouse2008} and demonstrate the increased accuracy of DMC calculations for the ground state energy of a C$_2$ molecule when using an MSJ trial wave function with optimized orbitals.
Both of these approaches to optimization maintain the orthogonality of determinants in the multi-Slater expansion.
On the other hand, results from multi-configuration calculations on some test atoms and molecules indicate that full-CI calculation accuracy can be achieved using wave functions with fewer optimized non-orthogonal determinants than if the determinants are kept orthogonal. \cite{Koch1993, McClean2015, Goto2013}
Optimized non-orthogonal Slater determinant expansions without a Jastrow have been used to accurately calculate bonding properties of molecules in valence bond theory \cite{doi:10.1021/ar50071a002}, ground state correlations of small molecules using Hartree-Fock with symmetry-projected wave functions \cite{doi:10.1063/1.4832476, doi:10.1080/00268976.2013.874623, PhysRevB.89.125129}, and correlation energies on the 1-d Hubbard model using resonanting Hartree-Fock with spin-projected wave functions. \cite{PhysRevB.69.045110, doi:10.1143/JPSJ.62.1653}
However, so far non-orthogonal determinants with a Jastrow have not been extensively explored as trial wave functions for FN-DMC calculations.

{In this paper, we investigate MSJ trial wave functions with non-orthogonal determinants for FN-DMC calculations, in which identical orbitals in different determinants are optimized independently.}
We conduct FN-DMC calculations for the ground state energy of a C$_2$ molecule on the first-principles Hamiltonian with electronic core potentials (ECPs) using three kinds of trial wave functions: MSJ states, MSJ states with optimized orthogonal determinants (MSJ+O), and MSJ states with optimized non-orthogonal determinants (MSJ+NO).
We used the QWalk\cite{Wagner2009} package for the QMC calculations, in which we implemented orthogonal and non-orthogonal optimization for this project. 
We find that the MSJ+NO trial wave function yields improvements to the ground state FN-DMC energies and single particle properties in addition to the improvements from using the MSJ+O trial wave function.

\section{Methods}
The first-principles Hamiltonian takes the form:
\begin{equation}
\hat{H} = -\frac{1}{2} \sum_i \nabla_i^2 + \sum_{i<j} \frac{1}{r_{ij}} + \sum_{i\alpha} \frac{Z_\alpha}{r_{i\alpha}} + \sum_{\alpha<\beta} \frac{Z_{\alpha}Z_{\beta}}{r_{\alpha \beta}}
\end{equation}
where $i,j$ are electron indices, and $\alpha$ is a nuclear index, and we have used atomic units. 
We employ a clamped nucleus approximation. 
ECPs from Burkatski \textit{et al} \cite{Burkatzki2007, Burkatzki2008} were used to eliminate the core electrons and give the form for $V_{\alpha}$.
We considered three different wave function \textit{ansatzes}, also called trial wave functions, summarized in Table~\ref{tab1}.

\begin{table}
  \caption{\label{tab1} Wave function \textit{ansatzes} and their corresponding variational parameters.}
\begin{tabular}{|c | c |c|} 
\hline
Name & Form & Parameters \\ 
\hline
MSJ & $\psi=e^{J(\vec{\alpha})}\sum_I C_I |D_I (\{ \phi\})\rangle$ & $\vec{\alpha}, \vec{C}$ \\
MSJ+O & $\psi=e^{J(\vec{\alpha})} e^{\Theta(\vec{\theta})}\sum_I C_I |D_I (\{ \phi\})\rangle$ & $\vec{\alpha}, \vec{C}, \vec{\theta}$\\
MSJ+NO & $\psi=e^{J(\vec{\alpha})} \sum_I e^{\Theta(\vec{\theta_I})} C_I |D_I (\{ \phi\})\rangle$ &$\vec{\alpha}, \vec{C},\{\vec{\theta_I}\}$\\
\hline
\end{tabular}
\end{table}
\begin{figure}
\includegraphics[width=0.40\textwidth]{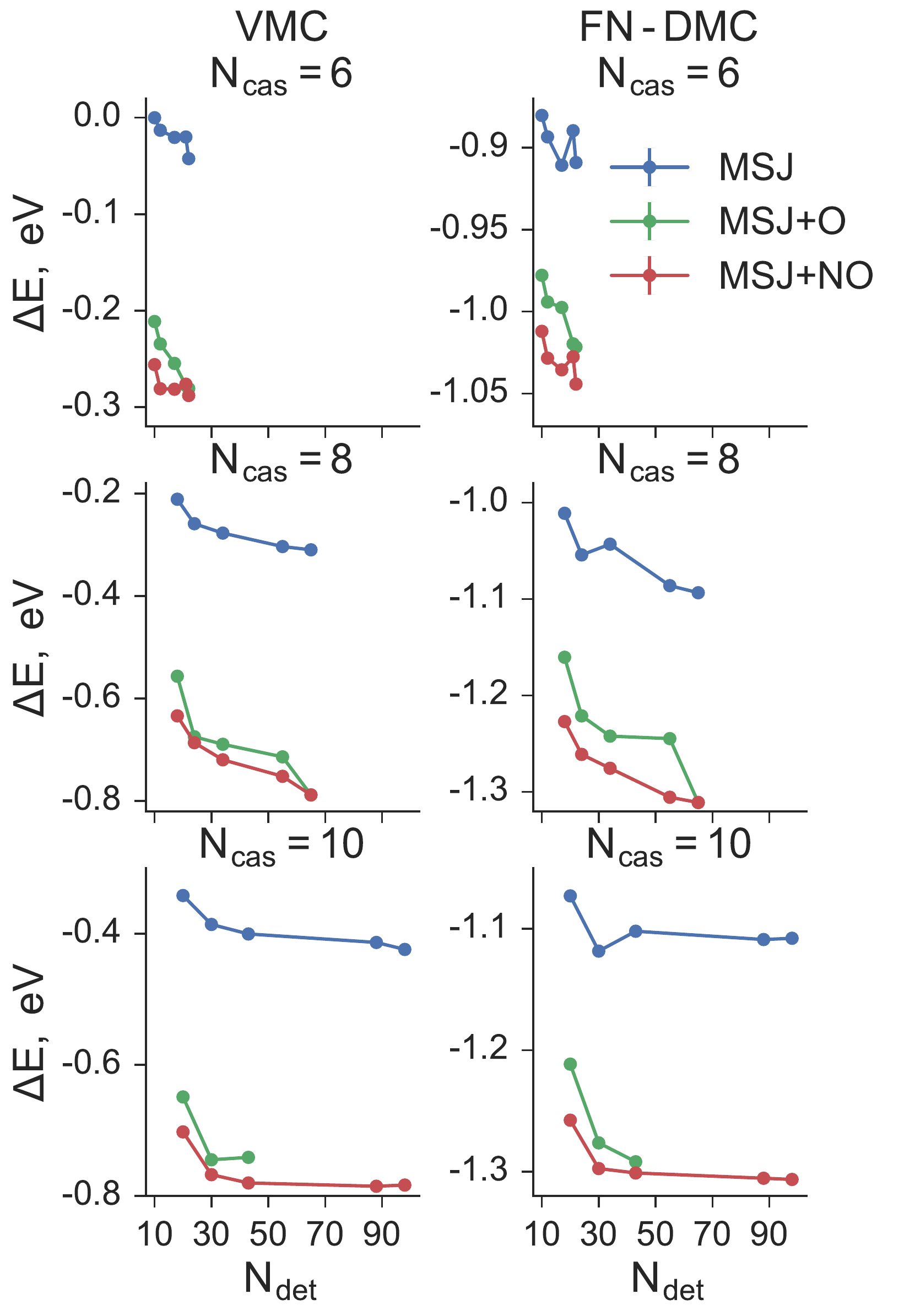}
\caption{\label{fig1} Optimized VMC energies and FN-DMC energies for trial wave functions with differing active space sizes (N\textsubscript{cas}) and number of determinants (N\textsubscript{det}). All energies are presented relative to the optimized VMC energy of the MSJ state with N\textsubscript{cas}=6, N\textsubscript{det}=10; $\Delta E=E - E_{MSJ}^{VMC}[$N\textsubscript{cas}=6,N\textsubscript{det}=10]. For each choice of N\textsubscript{cas}, N\textsubscript{det} there are three trial functions, MSJ, MSJ+O and MSJ+NO with sequentially lower VMC/FN-DMC energies. }
\end{figure}

{The simplest trial wave functions used are of MSJ form with optimized Jastrow and determinant coefficients}:
\begin{equation}
  \psi_{MSJ}(\vec{\alpha},\vec{C})=e^{J(\vec{\alpha})}\sum_I C_I |D_I (\{ \phi\})\rangle.
\end{equation} 
Here $e^J$ represents the Jastrow factor which depends on Jastrow parameters $\alpha$ followed by a multi-Slater determinant expansion with coefficients $C$ composed of single particle orbitals $\phi$, in this case from a restricted open Hartree-Fock calculation.
We constructed fifteen different MSJ states by attaching three-body Jastrow factors given in Wagner and Mitas \cite{Wagner2009} to distinct multi-Slater states calculated in PySCF. \cite{Sun2018}
These determinant expansions were constructed using a complete active space method with four active electrons per carbon atom and an active space $N_\text{cas}$ of 6,8, and 10.
Determinants were selected with coefficient weights $|C_I|$ greater than 0.05, 0.025, 0.0175, 0.01, and 0.0075, after which we saw little improvement in the total energy. 
This procedure generated 15 starting MSJ states, whose energies are reported in Fig~\ref{fig1} as MSJ.

{The MSJ+O trial wave functions build upon the MSJ trial wave functions by including orthogonal orbital rotations, the simplest form of orbital relaxation we will consider. }
To each of the fifteen MSJ states we now additionally apply an orbital rotation operator following Toulouse \textit{et al}\cite{Toulouse2008}, 
\begin{equation}
\psi_{MSJ+O}(\vec{\alpha},\vec{C},\vec{\theta})=e^{J(\vec{\alpha})} e^{\Theta(\vec{\theta})}\sum_I C_I |D_I (\{ \phi\})\rangle.
\end{equation}
The rotation operator $e^{\Theta(\vec{\theta})}$ acts on occupied orbitals as a rotation in orbital space 
\begin{equation}
\phi_i \rightarrow \sum_j [e^{\Theta(\vec{\theta})}]_{ij} \phi_j
\end{equation}
where the set $\{\phi_j\}$ is composed of select single particle orbitals that have the same symmetry as the occupied orbital $\phi_i$.
We included in the set $\{\phi_j\}$ all occupied orbitals with the same symmetry as orbital $\phi_i$ and the lowest energy unoccupied orbital with the same symmetry.
The rotation operator acts on different spin channels independently, therefore the optimized state may be spin contaminated: it may not be an eigenstate of the total $S^2$ operator.
Since the rotation operator acts identically on each determinant in a given spin channel, the determinants will remain orthogonal after optimization, so we refer to such a state as an MSJ state with optimized orthogonal determinants (MSJ+O state). 

{The MSJ+NO trial wave functions build upon the MSJ+O trial wave functions, allowing for non-orthogonal orbital rotations.}
We relax the orthogonality of determinants by moving the orbital rotation operator within the sum, such that 
\begin{equation}
\psi_\text{MSJ+NO}(\vec{\alpha},\vec{C},\{\vec{\theta}_I\})=e^{J(\vec{\alpha})} \sum_I e^{\Theta(\vec{\theta_I})} C_I |D_I (\{ \phi\})\rangle.
\end{equation}
Unlike the MSJ+O case, MSJ+NO \textit{ansatz} allows orbitals in different determinants to be optimized independently, potentially allowing for a more compact wave function.

For each trial wave function, we optimized the parameters listed in Table~\ref{tab1} using the linear method described in Toulouse \textit{et al}.\cite{Toulouse2007}
Filippi \textit{et al} describe an efficient method for calculating, in quantum Monte Carlo, the parameter derivatives required for the optimization. \cite{Filippi2016}. These optimized trial wave functions were then used in FN-DMC calculations. Diffusion Monte Carlo (DMC) is a quantum Monte Carlo method which projects out the ground state of a Hamiltonian given some initial trial wave function. 
Consider a trial wave function $|\psi\rangle$ and a Hamiltonian $H$ with ground state $|\phi_0\rangle$.
We apply the projector $e^{-\tau H}$ as $\tau \rightarrow \infty$ to $|\psi \rangle$
\begin{equation}
\lim_{\tau \rightarrow \infty} e^{-\tau H} |\psi\rangle \propto \langle \phi_0|\psi\rangle |\phi_0\rangle,
\end{equation}
projecting out the ground state $|\phi_0\rangle$ of $H$ as long as the trial wave function we choose is not orthogonal to the ground state.
The stochastic implementation of this projection on a many-body Hamiltonian with fermions leads to a fermion sign problem.
We deal with this sign problem through a fixed-node approximation, where the nodal surface of the projected wave function is forced to match that of the initial trial wave function.
This approximation makes FN-DMC variational, and will only return the exact ground state of $H$ if the nodal surfaces of $|\psi\rangle$ and $|\phi_0\rangle$ are identical.
We performed FN-DMC using T-moves\cite{PhysRevB.74.161102} with a timestep of $\tau = 0.01$. 
This choice of time step leads to a time step error of approximately 1mHa, which is an order of magnitude smaller than the differences in FN-DMC energies we are concerned with. 

Alongside energies, we also calculated VMC and FN-DMC single particle densities. 
In FN-DMC there is a systematic mixed-estimator error that affects the calculation of all quantities which do not commute with $H$, such as the charge density. 
In order to account for the first order of this mixed-estimator bias we extrapolate the FN-DMC single particle densities using the formula $\rho_{extrap} = 2\rho_{FN-DMC} - \rho_{VMC}$. \cite{Foulkes2001}
This bias does not appear in VMC.

\section{Results} 
Fig.~\ref{fig1} shows the variational Monte Carlo (VMC) and FN-DMC energies for the multiple trial wave functions described earlier. 
The VMC energies were calculated after the optimization of the variational parameters in the trial wave functions.
The method for calculation and N\textsubscript{cas} are shown in the subtitles. 
The average improvement in variational energy when using optimized orthogonal determinants is $\langle E_\text{MSJ} - E_\text{MSJ+O} \rangle = $0.32 eV with a standard deviation of 0.08 eV, where the average $\langle \rangle$ is over states at all N\textsubscript{cas}, N\textsubscript{det}.
The FN-DMC average improvement is smaller, at 0.14 eV, with a standard deviation of 0.03 eV.
The significant improvement in FN-DMC energy resulting from orthogonally optimizing determinants in trial wave functions indicates the importance of orbital relaxation for accurate QMC calculations.
\begin{figure}
\centering
\begin{minipage}[b]{0.2\textwidth}
\hspace*{-1.5cm}\includegraphics[height=0.5\textheight]{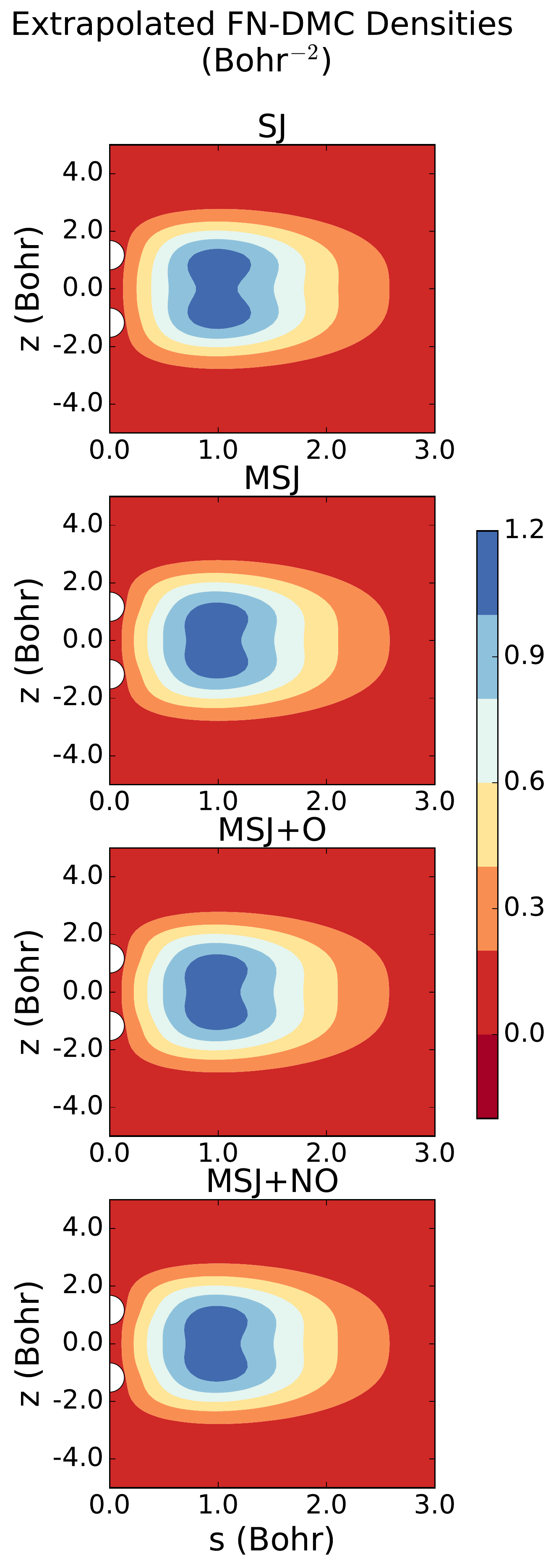}
\end{minipage}
\begin{minipage}[b]{0.2\textwidth}
\hspace*{-0.25cm}\includegraphics[height=0.4\textheight]{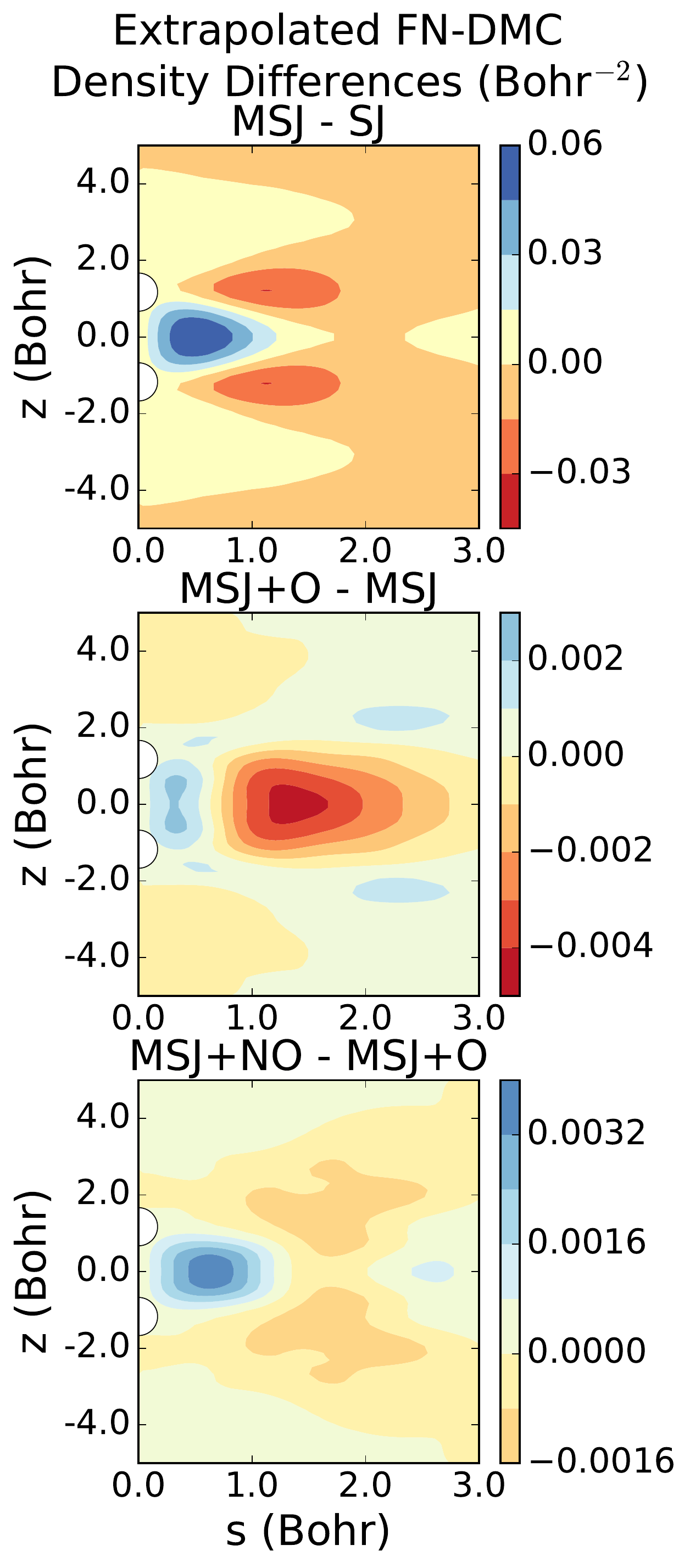}
\end{minipage}
\caption{\label{fig2} The left column shows cylindrically averaged ($\int d\phi \, \rho(s,\phi,z)\, 2\pi s$) FN-DMC charge densities for various optimized trial wave functions, with all multi-Slater trial functions built with N\textsubscript{cas}=10, N\textsubscript{det}=43. Here $s$ is the radial coordinate, $z$ is the z coordinate, and the cylindrical averaging was done over the azimuthal angle $\phi$. The right column shows differences in the charge densities in the left column.}
\end{figure}

{MSJ+NO trial wave functions provide additional improvements to the ground state energies.}
The average improvement in VMC energy is $\langle E_\text{MSJ+O} - E_\text{MSJ+NO} \rangle =$ 0.031eV (standard deviation 0.022 eV) and 0.032 eV (standard deviation 0.019 eV) for the FN-DMC energy. 
We believe the large standard deviation may arise from two main sources. 
First, different trial functions will have different changes in energy due to the non-orthogonal orbital rotation.
For example, closer to a complete multi-determinant expansion of the ground state, (N\textsubscript{det}, N\textsubscript{cas} $\rightarrow \infty$), the decrease in energy due to non-orthogonal orbital optimization will approach zero.
On the other hand, some finite multi-determinant expansion will have a non-zero decrease in energy due to this optimization.
Second, since we are now allowing for independent rotation parameters for each determinant, the variational space becomes N\textsubscript{det} times more complex than the space for orthogonal orbital optimization. 
A more complex variational space means that the linear optimization may more frequently get caught in local minima.
This effect does not appear in our calculations given the nearly monotonic decrease in VMC and FN-DMC energies as we increase N\textsubscript{det} and N\textsubscript{cas}. 

Further, the MSJ+NO trial wave functions can achieve similar FN-DMC energies with fewer determinants when compared with MSJ+O trial wave functions. For example, the FN-DMC energy when using an MSJ+NO trial wave function with 24 determinants is lower than when using an MSJ+O trial wave function with 55 determinants with active space size N\textsubscript{cas}=8. The introduction of optimized non-orthogonal determinants can therefore increase the compactness of QMC trial wave functions without sacrificing accuracy. 

To understand the effect of the improved trial functions, we compared the one particle density for each wave function type.
The first column of plots in Fig~\ref{fig2} shows the cylindrically averaged charge densities calculated in FN-DMC with mixed-estimator extrapolation for a Slater-Jastrow and MSJ, MSJ+O and MSJ+NO trial functions with N\textsubscript{cas}=10, N\textsubscript{det}=43. 
The second column in Fig~\ref{fig2} presents differences in these charge densities. 
The white semicircles represent the C atoms. 

Going from from the SJ to MSJ trial function, the extrapolated charge density increases between the two carbon atoms. 
This increase in bonding character makes sense since the ground state of C$_2$ is multi-reference in character. \cite{doi:10.1063/1.437480}
Orbital optimization continues this trend; however, the subsequent increase in charge density in the bonding region is an order of magnitude smaller than the increase between SJ and MSJ. 
This result is reasonable, since introducing correlation into trial wave functions allows for electrons to avoid each other while still occupying the bonding region.
The extrapolated spin densities exhibit spin contamination for the MSJ, MSJ+O and MSJ+NO trial functions, but the magnitude of the spin contamination is O(10$^{-4}$) Bohr$^{-2}$, an order of magnitude smaller than the charge redistribution. 

\section{Conclusion}
{We find that using MSJ+NO trial wave functions yield improvements to the ground state FN-DMC energy and single particle properties of a C$_2$ molecule in addition to the improvements from using MSJ+O trial wave functions.} 
For example, the FN-DMC energy calculated using an MSJ+NO trial function with only 24 determinants is lower than the FN-DMC energy using an MSJ+O trial function with 55 determinants.
Compared to an average decrease in the FN-DMC energy of 0.14 eV (standard deviation 0.03 eV) when using MSJ+O trial wave functions, using MSJ+NO trial wave functions provide an additional average reduction of 0.032 eV (standard deviation 0.019 eV).
Using trial wave functions with either orthogonally or non-orthogonally optimized determinants increases the bonding character of the FN-DMC charge density when compared to the density using a bare MSJ trial wave function.
Our results indicate that using non-orthogonal determinants in multi-Slater expansions may lead to more compact multi-Slater-Jastrow trial wave functions for small molecules.

\begin{acknowledgments}
This research is part of the Blue Waters sustained-petascale computing project, which is supported by the National Science Foundation (awards OCI-0725070 and ACI-1238993) and the state of Illinois. Blue Waters is a joint effort of the University of Illinois at Urbana-Champaign and its National Center for Superconducting Applications.
This work was funded by the grant DOE FG02-12ER46875 (SciDAC).
\end{acknowledgments}

\bibliography{biblio}

\end{document}